\documentclass[preprintnumbers, prd, twocolumn, showpacs, floatfix, preprintnumbers, letterpaper, nofootinbib, amsmath, amssymb, superscriptaddress]{revtex4}

\usepackage{amssymb, amsmath, bm, dcolumn, epsf, graphicx, latexsym, mathbbol, slashed, simplewick}

\usepackage{color}
\usepackage{ulem}
\newcommand{\tc}{\textcolor{black}}

\newcommand{\be}{\begin{equation}}
\newcommand{\ee}{\end{equation}}
\newcommand{\bea}{\begin{eqnarray}}
\newcommand{\eea}{\end{eqnarray}}
\newcommand{\barr}{\begin{array}}
\newcommand{\earr}{\end{array}}
\newcommand{\z}[1]{\zeta({\bf k}_{#1})}

\def\bk{{\bf k}}

\begin{document}

\title{Planck Constraints on Higgs Modulated Reheating of RG Improved Inflation }

\author{Yi-Fu Cai\footnote{Email: yifucai@physics.mcgill.ca}}
\affiliation{Department of Physics, McGill University, Montr\'eal, QC H3A 2T8, Canada}

\author{Yu-Chiao Chang\footnote{Email: f95222075@ntu.edu.tw}}
\affiliation{Department of Physics \& Leung Center for Cosmology and Particle Astrophysics, National Taiwan University, Taipei, Taiwan 10617}

\author{Pisin Chen\footnote{Email: chen@slac.stanford.edu}}
\affiliation{Department of Physics \& Leung Center for Cosmology and Particle Astrophysics, National Taiwan University, Taipei, Taiwan 10617}
\affiliation{Graduate Institute of Astrophysics, National Taiwan University, Taipei, Taiwan 10617}
\affiliation{Kavli Institute for Particle Astrophysics and Cosmology, SLAC, \\ Stanford University, Stanford, CA 94305, USA}

\author{Damien A. Easson\footnote{Email: easson@asu.edu}}
\affiliation{Department of Physics, Arizona State University, Tempe, AZ 85287, USA}

\author{Taotao Qiu\footnote{Email: qiutt@ntu.edu.tw}}
\affiliation{Department of Physics \& Leung Center for Cosmology and Particle Astrophysics, National Taiwan University, Taipei, Taiwan 10617}


\begin{abstract}
Within the framework of RG improved inflationary cosmology motivated by {\it{asymptotically safe gravity}}, we study the dynamics of a scalar field which can be interpreted as the Higgs field. The background trajectories of this model can provide sufficient inflationary e-folds and a graceful exit to a radiation dominated phase. We study the possibility of generating primordial curvature perturbations through the Standard Model Higgs boson. This can be achieved under finely tuned parameter choices by making use of the modulated reheating mechanism. The primordial non-gaussianity is expected to be sizable in this model. Though tightly constrained by the newly released Planck CMB data, this model provides a potentially interesting connection between collider  and early universe physics.
\end{abstract}

\maketitle

\section{Introduction}

Inflation is a successful model for the early universe, and manages to address several challenges of Big Bang cosmology \cite{Guth:1980zm, Linde:1981mu, Albrecht:1982wi} (see \cite{Starobinsky:1980te, Fang:1980wi, Sato:1980yn} for early works). One of inflation's crowning achievements is the prediction of a nearly scale-invariant primordial power spectrum which was verified to high precision by Cosmic Microwave Background (CMB) observations \cite{Komatsu:2010fb, Ade:2013lta, Ade:2013tta}. The success of inflation, however, is based on a series of assumptions including the existence of a sufficiently long period of quasi-exponential expansion realized by an as yet unobserved slow-roll scalar field. An important question to be addressed is whether such a scalar field with an appropriate potential indeed exists in nature.

The Higgs boson is a scalar field predicted by the Standard Model (SM) of particle physics. Recent results from the ATLAS and CMS collaborations at CERN's Large Hadron Collider (LHC) have established the existence of a new resonance at $125$ GeV, which is compatible with the Higgs boson with $5\sigma$ significance for the signal \cite{ATLAS:2012, CMS:2012}. As the only existing scalar particle in the SM, it is natural to speculate that the Higgs field may be used to play a dual role as the inflaton. However, the corresponding Higgs energy scale is much too low compared with the typical inflationary scale. In order to solve this problem, a non-minimal coupling was introduced between the Ricci scalar and the Higgs field \cite{Bezrukov:2007ep, DeSimone:2008ei} (see also an earlier attempt in \cite{CervantesCota:1995tz}), and the associated asymptotic freedom was studied in \cite{Barvinsky:2009fy}. It was soon realized that such a non-minimal coupling leads to a unitarity violation energy scale that must be larger than the inflationary scale; otherwise the effective field description would fail (for example, see \cite{Burgess:2009ea, Barbon:2009ya, Lerner:2009na, Burgess:2010zq, Bezrukov:2010jz, Atkins:2010yg, Horvat:2011wr} for detailed discussions). One thus faces challenges when constructing a suitable inflation model based on the Higgs field.

One of the most challenging tasks in physics is the construction of a consistent UV-complete theory of gravity. Weinberg made an intriguing proposal in the 1970's: that the effective description of quantum gravity may be non-perturbatively renormalizable via the notion of asymptotic safety (AS) \cite{Weinberg:1977, Weinberg:1979}. In such a scenario the flows of the renormalization group (RG) approach a fixed point in the ultraviolet (UV) limit, and a finite dimensional critical surface of trajectories evolves to this point at short distance scales \cite{Weinberg:2009bg, Weinberg:2009wa}. Such a fixed point can be found in the Einstein-Hilbert truncation \cite{Reuter:1996cp}, and the AS scenario has been studied extensively in the literature \cite{Souma:1999at, Bonanno:2001xi, Lauscher:2001ya, Litim:2003vp, Codello:2006in, Cai:2010zh} (for recent reviews see \cite{Niedermaier:2006ns, Percacci:2007sz}).

Recently, two of the present authors (YFC and DAE) proposed a cosmological model based on the concept of asymptotically safe gravity, in which the Higgs boson plays an important role during inflation \cite{Cai:2012qi}. While there is as yet no explicit proof of the AS behavior, with its theoretical promise it warrants to explore whether the gravity theories under such a scenario are experimentally testable. It has been argued that if there are no intermediate energy scales between the SM and AS scales, the mass of the Higgs boson is predicted to be $m_H=126 \,{\rm GeV}$, with only several {\rm GeV} uncertainty \cite{Shaposhnikov:2009pv}. Inspired by this idea, in our model we consider the gravitational constant $G$ and the cosmological constant $\Lambda$ varying with the energy scale, and thus varying throughout the cosmological evolution. Since the gravitational constant and the cosmological constant vary along with the running of the cutoff scale, the stress-energy tensor acquires an extra contribution. When taking into account the extra term, the model is found to correspond to a $f(R)$ model \cite{Cai:2011kd} (see also \cite{Bonanno:2012jy, Hindmarsh:2012rc} for generalized discussions). As a consequence, there are effectively two scalar degrees of freedom, one being the adiabatic mode and the other being an iso-curvature mode. We find that the corresponding perturbation theory leads to primordial power spectra for both the curvature perturbation and the entropy perturbation. When the cutoff scale runs lower than certain critical value, the universe exits inflation gracefully and the AS gravity approaches GR simultaneously.

After inflation, there exist two possible mechanisms of converting entropy fluctuations into curvature perturbations. The first is to make use of the curvaton mechanism \cite{Mollerach:1989hu, Linde:1996gt, Enqvist:2001zp, Lyth:2001nq, Moroi:2001ct} . However, in the model of AS inflation such a transfer is highly nonlinear and thus leads to severe fine-tuning problem on model parameters \cite{Cai:2012qi} (see also \cite{Kunimitsu:2012xx, Choi:2012cp} for Higgs-curvaton problems in GR). Recently, a second possibility of realizing such a transfer was suggested in Refs.\cite{DeSimone:2012qr, DeSimone:2012gq}, where the Higgs boson could modulate the efficiency of reheating\cite{Dvali:2003em, Dvali:2003ar, Kofman:2003nx}. This mechanism requires that the decay rate of the inflaton to be a function of a second field with a much lighter mass. This allows the entropy fluctuations during inflation to transfer to curvature perturbations on a hypersurface with the average Hubble expansion comparable to the inflaton decay rate.

In the present paper we study the possibility of AS inflation with modulated reheating through the Higgs boson. The paper is organized as follows. Section II gives a brief review on the AS scenario of gravity theory with Einstein truncation and its classical correspondence to $f(R)$ gravity is introduced. In Section III, we perform a conformal transformation from Jordan frame to Einstein frame and then study the inflationary background driven by the RG improved gravitational and cosmological constants. We show that this cosmological system is equivalent to a double field inflation model where the contribution of the Higgs boson is secondary. Section IV is devoted to the analysis of modulated reheating via the Higgs field during which the isocurvature fluctuation is converted into the adiabatic perturbation and thus forms the primordial power spectrum of curvature perturbation. The primordial non-gaussianity seeded by nonlinear fluctuations of the Higgs is computed as well. In Section V, we confront our model with the newly released Planck data and find that the parameter space is tightly constrained. Finally, we conclude with a discussion in Section VI. We will work with the reduced Planck mass, $M_{p} = 1/\sqrt{8\pi G_N}$, where $G_N$ is the gravitational constant in the IR limit, and adopt the mostly-plus metric sign convention $(-,+,+,+)$.

\section{Asymptotically safe gravity}

We begin with a brief review of the cosmological of AS gravity minimally coupled to ordinary matter Lagrangian. Specifically, we consider AS gravity with  Einstein-Hilbert truncation and corresponding beta functions including next-to-leading order corrections of gravitational coupling. This type of gravitational theory can be reformulated as an $f(R)$ theory non-minimally coupled to the matter Lagrangian.

\subsection{Asymptotic safety}

We start with a RG inspired effective gravitational Lagrangian with Einstein-Hilbert truncation,
\begin{eqnarray}
 {\cal L}_{\rm AS} = \frac{R-2\Lambda(p)}{16\pi G(p)} ~,
\end{eqnarray}
where $p$ is the RG cutoff scale beyond which the UV modes are argued to be integrated out. This effective Lagrangian automatically connects with ordinary Einstein gravity in the IR regime where the gravitational and cosmological constants flow to some integral constants that are constrained by observations. In the UV limit, these ``constants" flow to a UV fixed point according to their beta functions. Quantum corrections are therefore encoded in the evolution of the coupling constants as functions of the cutoff scale, whose beta functions can be extracted from the RG equations.

We define the dimensionless gravitational and cosmological constants as follows,
\begin{eqnarray}
 g(p) \equiv \frac{p^2}{24\pi}G(p)~,~~\lambda(p) \equiv \frac{\Lambda(p)}{p^2}~.
\end{eqnarray}
Given the exact forms of RG equations, one can follow the flows of $g$ and $\lambda$ along with the cutoff scale and obtain a fixed point in the UV limit. The AS scenario suggests that this UV fixed point is  attractive. Note that the explicit forms of beta-functions depend on the choice of the cutoff function and the relevant gauges. In Ref. \cite{Weinberg:2009wa}, it was observed that the UV fixed point often corresponds to a de Sitter solution, however neither the energy scale of the background nor the amplitude of quantum fluctuations provide a successful application to early universe inflationary cosmology\cite{Cai:2011kd}.

If the RG-improved gravity theory is viable, its RG trajectory should connect smoothly with standard Einstein gravity in the IR limit so as to be consistent with astronomical and cosmological observations. We would therefore like to study the RG improved gravity theory in the regime that is sufficiently close to GR while still retaining linearized quantum corrections to the beta-functions. To begin the analysis, we linearize the beta-functions for dimensionless coupling constants as follows:
\begin{eqnarray}
 \beta_\lambda &\equiv& p\partial_p\lambda = -2\lambda+2\alpha g~, \\
 \beta_g &\equiv& p\partial_pg = 2g -2\beta^2 g^2/3~,
\end{eqnarray}
which include next-to-leading order corrections to $g$. The coefficients $\alpha$ and $\beta$ are cutoff functions dependent. Under this parameterization, one can obtain approximate forms of the dimensionless couplings, which are given by
\begin{eqnarray}
 g(p) &\simeq& \frac{3 G_N p^2}{72\pi +\beta^2 G_N p^2}~,\\
 \lambda(p) &\simeq& \frac{\Lambda_{IR}}{p^2} +\frac{3\alpha}{\beta^2}
 -\frac{216\pi\alpha}{\beta^4G_Np^2} \ln[\frac{72\pi}{G_N} +\beta^2 p^2]~,
\end{eqnarray}
where $G_N$ and $\Lambda_{IR}$ are Newton's constant and the cosmological constant in the infrared limit and thus correspond to those in GR. The above two couplings approach non-vanishing constant values in the $p\ll M_p$ limit and therefore have the expected AS behavior. Note also that the parameters $\alpha$ and $beta$ should in principle be calculated from concrete theory of quantum gravity rather than free model parameters. However, because it is not know how to do this calculation, one can have many different possibilities and thus can treat them as free parameters effectively. The corresponding parameter choice can determine the way $g$ and $\lambda$ approach to their fixed point, and thus could impose possible constraint on the theory. As will be seen later, we choose the values of $\alpha$ and $\beta$ such that our model can fit the observational data.

The corresponding RG improved gravitational and cosmological constants obey the following relations,
\begin{eqnarray}
 \label{G_AS}
 G &\simeq& \frac{G_N}{1+\xi_G G_Np^2}~, \\
 \label{Lambda_AS}
  \Lambda &\simeq& \Lambda_{IR} +\xi_\Lambda p^2 -\frac{\xi_\Lambda}{\xi_G} G_N^{-1} \ln[1+\xi_GG_Np^2] ~,
\end{eqnarray}
where $\xi_G$ and $\xi_\Lambda$ are the model parameters determined by RG flow coefficients through
\begin{eqnarray}
 \xi_G = \frac{\beta^2}{72\pi} ~~ {\rm and} ~~ \xi_\Lambda = \frac{3\alpha}{\beta^2}~.
\end{eqnarray}
When $p\rightarrow 0$, they approach the classical values determined by observations and thus GR is recovered in the IR limit. Conversely, in the extreme UV regime the value of $G$ approaches zero, which implies a weakly coupled gravitational system at extremely high energy scale. In between, we expect a period of sufficiently slow variation of $\Lambda$ and thus the occurrence of a inflationary phase at early times of cosmological evolution. We note that if $\xi_G$ is chosen to be much smaller than unity, one can Taylor expand the last term of Eq. \eqref{Lambda_AS} and the simplified expression
\begin{eqnarray}\label{Lambda_AS1}
 \Lambda \simeq \Lambda_{IR} + \frac{1}{2}\xi_\Lambda\xi_GG_Np^4~,
\end{eqnarray}
even if $p$ is of order $M_p$. We will find that this condition is necessary in order to achieve a viable inflationary phase.

\subsection{The $f(R)$ correspondence}

We consider minimal coupling between the AS gravity and the matter field,
\begin{eqnarray}
 S_{\rm AS} = \int d^4x \sqrt{-g} \left[ {\cal L}_{\rm AS} + {\cal L}_{\rm m} \right]~,
\end{eqnarray}
where ${\cal L}_{\rm m}$ is the Lagrangian of the matter field. As the gravitational constant varies along the cutoff scale $p$ which can be a function of space-time, varying the Lagrangian with respect to the metric yields the generalized Einstein equation
\begin{equation}
 R_{\mu\nu} - \frac{R}{2}g_{\mu\nu} + \Lambda g_{\mu\nu} = 8\pi GT^{(m)}_{\mu\nu} + G(\nabla_\mu\nabla_\nu-g_{\mu\nu}\Box)G^{-1} ~,
\end{equation}
where we have introduced the covariant derivative $\nabla_\mu$ and the operator $\Box \equiv g^{\mu\nu}\nabla_\mu\nabla_\nu$. Additionally, consistency of the Bianchi identity requires  the running of the cutoff scale to obey the constraint
\begin{eqnarray}\label{eom_p}
 \frac{R-2\Lambda(p)}{2G(p)}\nabla_\mu G(p) +\nabla_\mu \Lambda(p) = 0~.
\end{eqnarray}
The continuity equation of energy density  determines the dynamics of matter components and allows derivation of the equations of motion by varying the Lagrangian with respect to matter fields. The dynamics of this cosmological system are completely determined.

Inserting the forms of RG modified gravitational constant $G$ (Eq.\eqref{G_AS}), and cosmological constant $\Lambda$ (Eq.\eqref{Lambda_AS1}), into Eq.\eqref{eom_p}, one can identify the relation between the Ricci scalar and the cutoff scale,
\begin{eqnarray}
 p^2 \simeq \frac{R-2\Lambda_{IR}}{2\xi_\Lambda} - \frac{3\xi_GG_NR^2}{8\xi_\Lambda^2}~.
\end{eqnarray}
The original theory may be reformulated as an effective $f(R)$ model, where ${\cal L}_{\rm AS}$ is replaced by
\begin{eqnarray}
 f(R) &=& - \frac{\Lambda_{,p}}{8\pi G_{,p}}(R) \nonumber\\
  &\simeq& \frac{R-2\Lambda_{IR}}{16\pi G_N} +\frac{\xi_G}{32\pi\xi_\Lambda} (R-2\Lambda_{IR})^2~.
\end{eqnarray}
The subscript $,p$ denotes the derivative with respect to $p$. In general, the correspondence between the AS gravity and $f(R)$ theory holds if the Einstein-Hilbert truncation is applied. However, the detailed expression of $f(R)$ depends strongly on the specific forms of RG functions as well as the identification between the cutoff scale and Ricci scalar. We refer to \cite{Hindmarsh:2012rc} for a general discussion on this issue.

\section{$R^2$ Inflationary cosmology}

In a realistic cosmological model, the value of $\Lambda_{IR}$ is determined by observations pertaining to  late-time acceleration which are typically of the order $O(10^{-121})M_p^2$. Therefore, its contribution to the early universe background dynamics is totally negligible. For the time being, we neglect it and  our model reduces to a $R^2$ inflationary cosmology \cite{Starobinsky:1980te}.

In addition, for the matter field Lagrangian we focus on the SM Higgs scalar. We use the unitary gauge for the Higgs boson and temporarily neglect all gauge interactions. As a consequence, the Lagrangian of the matter field is given by
\begin{eqnarray}
 {\cal L}_m \supset -\frac{1}{2}\partial_\mu h\partial^\mu h - V(h) -V_{int} ~,
\end{eqnarray}
where $V(h)$ is the potential of the Higgs boson and $V_{int}$ represents the interactions between the Higgs and other particles in the standard model of particle physics. Without considering interactions with other particles, the form of the potential is approximately,
\begin{eqnarray}\label{V_h}
 V(h) \simeq \frac{\lambda}{4} (h^2-v^2)^2~,
\end{eqnarray}
in which $v$ is the vacuum expectation value of the Higgs boson,  with associated Higgs mass, $m_H =\sqrt{2\lambda}v$.

\subsection{Background equations of motion}

Now we turn our attention to inflationary solutions. It is convenient to perform a Weyl rescaling,
\begin{eqnarray}
 g_{\mu\nu}\rightarrow \tilde{g}_{\mu\nu} = \Omega^2g_{\mu\nu}~,
\end{eqnarray}
where $\Omega$ is the conformal factor defined by a newly introduced scalar field as follows,
\begin{eqnarray}\label{Omega}
 \Omega(\phi) = e^{ \frac{\phi}{\sqrt{6}M_p} }~.
\end{eqnarray}

As a result, the original AS system is equivalent to a two-scalar-field system minimally coupled to Einstein gravity without RG running, for which the effective Lagrangian is given by
\begin{eqnarray}
 {\cal L} \supset \frac{\tilde{R}}{16\pi G_N} - \frac{(\tilde\nabla\phi)^2}{2} - \frac{(\tilde\nabla{h})^2}{2\Omega^2(\phi)} - \tilde{V}(\phi, h)~,
\end{eqnarray}
with
\begin{eqnarray}
 \tilde{V}(\phi, h) =  U(\phi) + \frac{V(h)}{\Omega^4(\phi)} ~,
\end{eqnarray}
where the potential of the new scalar field takes the form:
\begin{eqnarray}
 U(\phi) = 2\pi M_p^4 \frac{\xi_\Lambda}{\xi_G} \left( 1-e^{-\frac{2\phi}{\sqrt{6}M_p}} \right)^2~.
\end{eqnarray}
This potential is sufficiently flat in the regime where $\phi \ll M_p$ and migrates into the quadratic form around $\phi=0$. This scalar field can thus play the role of the inflaton under a careful selection of RG running coefficients.

Substituting the flat Friedmann-Robertson-Walker (FRW) metric, $ds^2=-dt^2+a^2(t)d\vec{x}^2$, the Friedmann equation becomes
\begin{eqnarray}
\label{Friedman}
 {H}^2=\frac{1}{3M_p^2}\tilde{\rho}~,~\dot{{H}}=-\frac{1}{2M_p^2}(\tilde{\rho}+\tilde{P})~,
\end{eqnarray}
where $H\equiv\dot{a}/a$ and the dot denotes the time derivative in the Einstein frame. The energy density and the pressure in Einstein frame are:
\begin{eqnarray}
\label{rho}
 {\rho} &=& \frac{1}{2}\dot\phi^2 +\frac{1}{2\Omega^2(\phi)}{\dot h}^2 + U(\phi) + \frac{V(h)}{\Omega^4(\phi)}~,\\
\label{pressure}
 {P} &=& \frac{1}{2}\dot\phi^2 +\frac{1}{2\Omega^2(\phi)}{\dot h}^2- U(\phi) - \frac{V(h)}{\Omega^4(\phi)}~.
\end{eqnarray}
By varying the Lagrangian with respect to $\phi$ and $h$, we obtain the equations of motion for the scalar fields:
\begin{eqnarray}
\label{eom_phi}
 \ddot\phi +3{H}\dot\phi +U_{,\phi} -\frac{\Omega_{,\phi} }{\Omega^5}V +\frac{\Omega_{,\phi} }{\Omega^3} {\dot h}^2 &=& 0 ~,\\
\label{eom_h}
 \ddot h +3{H}\dot h -2\frac{\Omega_{,\phi} }{\Omega}\dot\phi\dot h +\frac{V_{,h}}{\Omega^2} &=& 0~.
\end{eqnarray}

\subsection{Slow roll inflation}

Eqs.\eqref{eom_phi} and \eqref{eom_h} reveal that the inflaton and the Higgs fields are coupled and the system is rather intricate. Fortunately, the coupling terms are dramatically suppressed during inflation due to the slow roll condition. We now introduce the series of slow roll parameters
\begin{eqnarray}
\label{epsilons}
 && \epsilon \equiv -\frac{\dot{H}}{H^2}~,~ \epsilon_\phi\equiv\frac{\dot\phi^2}{2M_p^2H^2}~,~ \epsilon_h\equiv\frac{\dot{h}^2}{2\Omega^2M_p^2H^2}~,\\
\label{etas}
 && \eta_\phi\equiv\frac{\tilde{V}_{,\phi\phi}}{3H^2} ~,~ \eta_h\equiv\frac{\tilde{V}_{,hh}}{3H^2} ~,~ \eta_{\phi h}\equiv\frac{\tilde{V}_{,\phi h}}{3H^2} ~,
\end{eqnarray}
for a cosmological system of coupled double fields. We note that the potential of $\phi$ becomes very flat when $\phi$ is larger than $M_p$ and in comparison the parameters $\epsilon_\phi$ and $\eta_\phi$ are relatively small. Simultaneously, other parameters are also very small due to the suppression by the large value of the conformal factor $\Omega$.

As a consequence, under the slow-roll approximation the background dynamics are determined by the following solutions:
\begin{eqnarray}
\label{sol_inf}
 \dot\phi \simeq -\frac{U_{,\phi}}{3H}~,~
 \dot{h} \simeq -\frac{V_{,h}}{3\Omega^2H}~,~
 H^2 \simeq \frac{U}{3M_p^2}~,
\end{eqnarray}
which implies a quasi-exponential expansion at early times. Since inflation ends when $\epsilon_\phi =1$, the substitution of the background solution for $\dot\phi$ in Eq.\eqref{sol_inf} into the slow-roll parameter $\epsilon_\phi$ in Eq.\eqref{epsilons} yields the value of $\phi$ at the end of inflation:
\begin{eqnarray}
 \phi_f \simeq \frac{\sqrt{6}}{2}M_p\ln(1+\frac{2}{\sqrt{3}})~.
\end{eqnarray}
As the contribution of the Higgs field during inflation is negligible, the number of e-folding of inflation mainly depends on $\phi$ through the relation ${\cal N}=-\int_i^f U d\phi/M_p^2U_{,\phi}$, which is given by
\begin{eqnarray}
 {\cal N}(\phi) \simeq \frac{3}{4}e^{\frac{2\phi}{\sqrt{6}M_p}}-\frac{3}{2}\frac{\phi}{\sqrt{6}M_p}-1.04~.
\end{eqnarray}
It is easy to achieve ${\cal N}=60$ if initially the inflaton is placed at $\phi_i\simeq 5.46 M_p$. Applying the slow-roll condition, one obtains the Hubble rate
\begin{eqnarray}
 H_I \simeq \sqrt{\frac{2\pi}{3}\frac{\xi_\Lambda}{\xi_G}}M_p~,
\end{eqnarray}
during inflation.

Moreover, the slow-roll parameters for the Higgs field $h$ are automatically small due to the suppression of the large value of the conformal factor $\Omega \sim O(10)$. Thus during inflation $h$ also varies slowly.
As is well known, the inflationary phase is an attractor solution in an expanding universe, and thus it is expected that other matter fields would be dominant in the pre-inflationary phase. Specifically, in our model the universe was dominated by the Higgs field in pre-inflationary era and at that moment the slow roll condition was not satisfied. As a result, the parameter $\eta_h$ can be larger or of order of unity. Therefore, we can make use of the relation $\eta_h \simeq 1$ to estimate the amplitude of the Higgs field at the initial moment of inflation, which requires, $\eta_h = V_{,hh}/3H^2\Omega^4 \lesssim 1$.
Therefore, one can estimate the initial amplitude of the Higgs at the beginning of inflation as
\begin{eqnarray}\label{h_i}
 h_i \simeq \frac{\Omega_i^2}{\sqrt{\lambda}}  H_I ~.
\end{eqnarray}
At the end of inflation there is no more suppression on slow-roll parameters of $h$ and their values become of the order of unity simultaneously. As a consequence, the amplitude of the Higgs field at the end of inflation is estimated as
\begin{eqnarray}\label{h_f}
 h_f \simeq \frac{2}{\sqrt{\lambda}} H_I~.
\end{eqnarray}
Combing Eqs.\eqref{h_i} \& \eqref{h_f}, we easily verify that during inflation the Higgs boson is required to satisfy the inequality:
\begin{eqnarray}
\label{h_I}
 h_f < h_I <h_i~.
\end{eqnarray}

Note that the Higgs field during inflation is not required to be less than the Hubble rate since its background energy density contributed is dramatically suppressed by the large value of the conformal factor. Our model therefore evades the theoretical constraint suggested in \cite{Choi:2012cp}. We will see that this is key to realizing Higgs-modulated reheating. The following numerics verify this result explicitly.

\begin{figure}[htbp]
\includegraphics[scale=0.45]{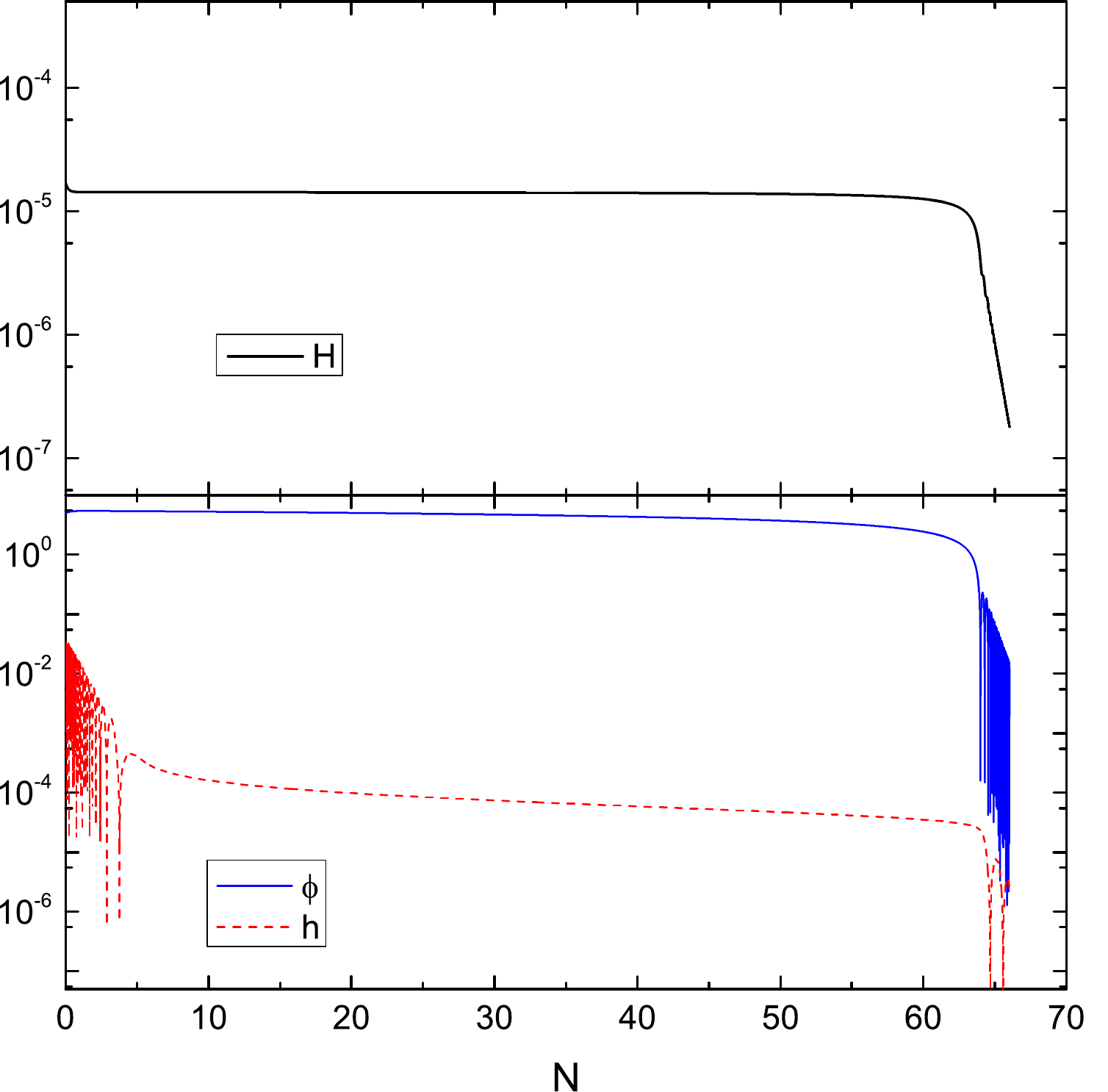}
\caption{Evolution of the Hubble parameter $H$ and two scalar fields $\phi$ and $h$ as functions of the e-folding number ${\cal N}$. In the solutions, the model parameters are, $\xi_G=0.72$ and $\xi_\Lambda=10^{-10}\xi_G$. The parameters of the potential for the Higgs are taken as, $\lambda=0.13$ and $v=246 {\rm GeV}$ according to particle physics observations. Initial field values are taken as $\phi_i = 5.46 M_p$ and $h_i = 10^{-2} M_p$. Planck units are adopted in the figure.}\label{Graph_bg}
\end{figure}

\begin{figure}[htbp]
\includegraphics[scale=0.5]{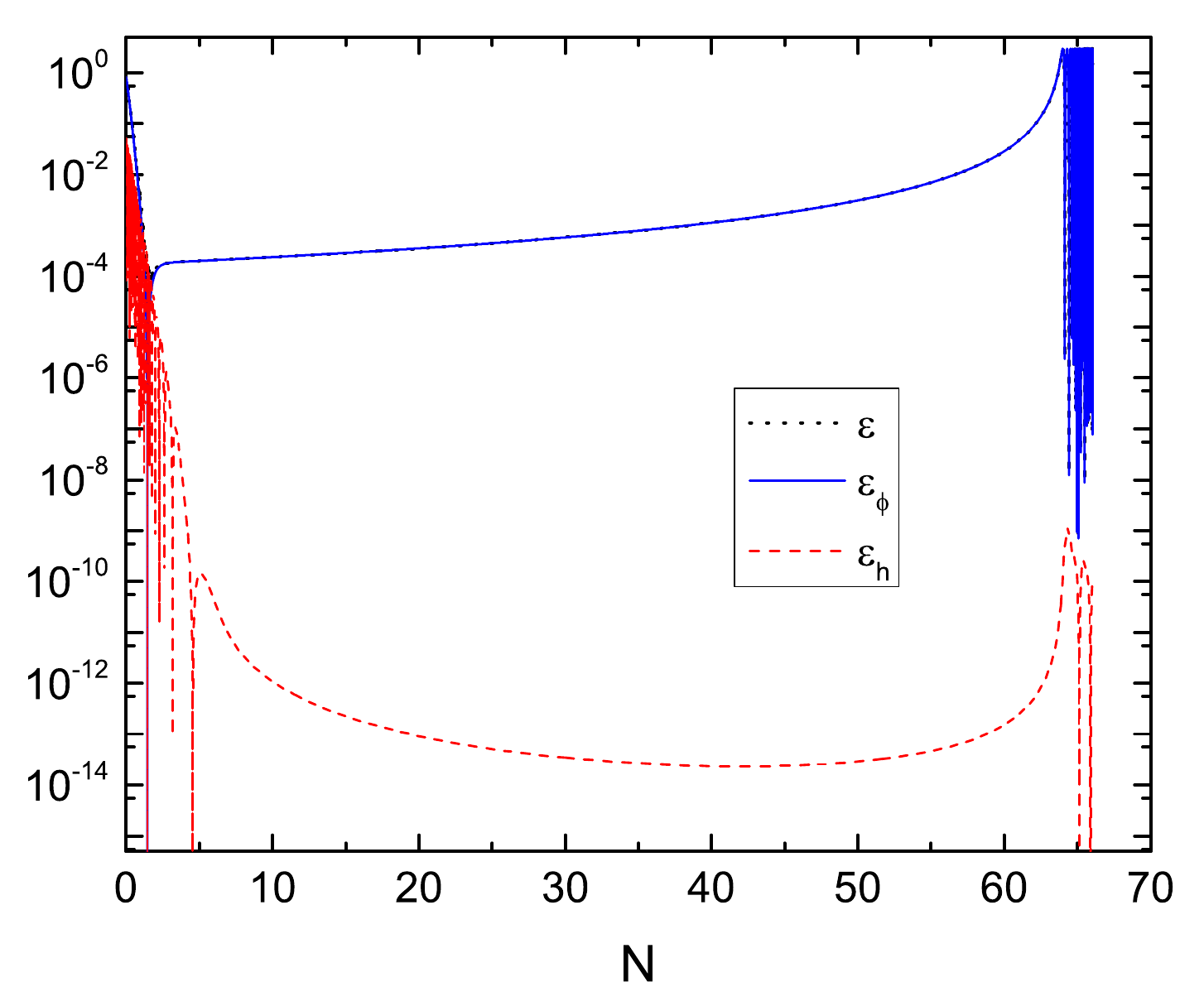}
\caption{Evolution of the slow-roll parameters $\epsilon$, $\epsilon_\phi$ and $\epsilon_h$ as functions of the e-folding number ${\cal N}$. The model parameters and initial conditions for this plot are the same as those for Fig.\ref{Graph_bg}. }
\label{Graph_epsilon}
\end{figure}

\begin{figure}[htbp]
\includegraphics[scale=0.5]{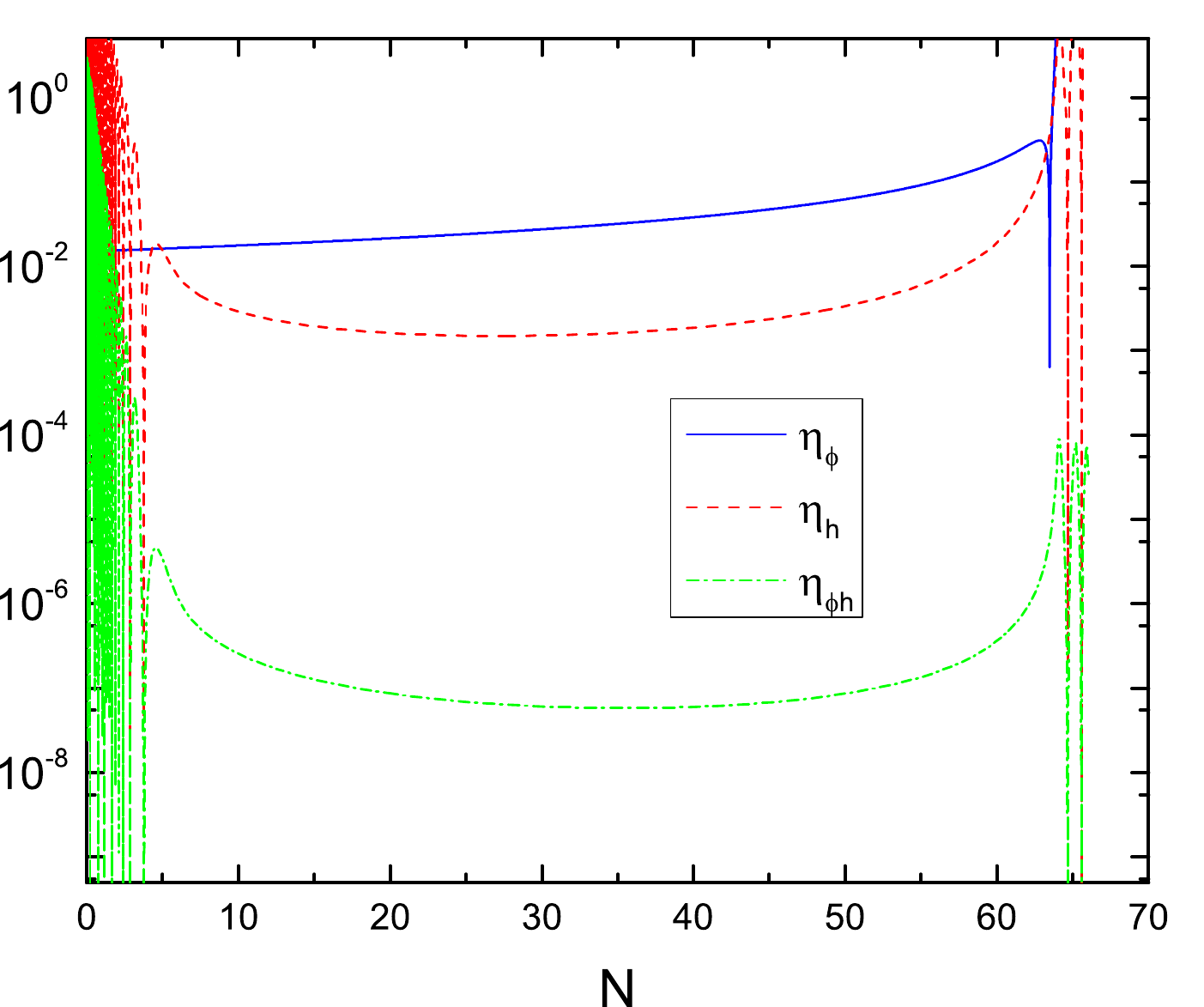}
\caption{Evolution of the slow roll parameters $\eta_\phi$, $\epsilon_h$ and $\epsilon_{\phi h}$ as functions of the e-folding number ${\cal N}$. The model parameters and initial conditions for this plot are the same as those for Fig.\ref{Graph_bg}. }
\label{Graph_eta}
\end{figure}

In the above we presented analytic solutions to inflationary dynamics. We now verify the results with  numerical computations. The results are shown in Figs.\ref{Graph_bg}, \ref{Graph_epsilon} and \ref{Graph_eta}.

Fig.\ref{Graph_bg} shows that the Hubble parameter varies very slowly in the middle region, which corresponds to the inflationary period. In contrast, the Higgs boson oscillates dramatically at the beginning of the evolution, which implies that the universe is dominated by the Higgs field in the pre-inflation phase. The transition from the Higgs dominated pre-inflation phase to the inflation phase follows an attractor solution that does not strongly depend on the choice of initial conditions. However, this result also implies that such a scenario has to meet the big bang singularity if one traces backwards in the cosmic evolution. From Fig.\ref{Graph_bg}, one can see that inflation ends when the value of $\phi$ decreases below $M_p$. The corresponding e-folding number is roughly $65$. Subsequently, the inflaton field  oscillates around $\phi=0$ which corresponds to an IR fixed point of AS gravity. Therefore, GR is automatically recovered at the end of inflation.

Fig.\ref{Graph_epsilon} and Fig.\ref{Graph_eta} show the evolutions of slow-roll parameters defined in Eq.\eqref{epsilons} and Eq.\eqref{etas} along with the cosmic expansion. In Fig.\ref{Graph_epsilon}, the background slow-roll parameter $\epsilon$ almost coincides with that for inflaton, $\epsilon_\phi$, since inflation is driven by the effects of RG-modified gravitational and cosmological constants. Among those parameters associated with the Higgs boson, the value of $\eta_h$ is the first to break the slow-roll condition after inflation. Consequently, the method of determining the value of $h$ during inflation by requiring $\epsilon_\phi\simeq 1$ is reliable. By substituting the parameter choices provided in Fig.\ref{Graph_bg} into the expression Eq.\eqref{h_I}, we find $h_I\simeq 10^{-4}$, which is consistent with the numerical result shown in the lower panel of Fig.\ref{Graph_bg}.

\subsection{Higgs dependent decay after inflation}

After the inflaton field rolls below the critical value $\phi_f$, it starts to oscillate around the IR point which corresponds to the GR limit. One can thus perform a Taylor expansion of the potential around $\phi=0$:
\begin{align}\label{V_osc}
 \tilde{V}(\phi, h) &\simeq \frac{1}{2}M_\phi^2\phi^2 +V(h) \nonumber\\
 & + V_{int}\left(1-\frac{4\phi}{\sqrt{6}M_p}+\frac{4\phi^2}{3M_p^2}\right) + O(\phi^3),
\end{align}
up to the $\phi^2$ order. In the above expression, we have introduced an effective mass for the inflaton defined by
\begin{eqnarray}\label{M_phi}
 M_\phi^2 \equiv \frac{2\pi}{3}M_p^2\frac{\xi_\Lambda}{\xi_G}~.
\end{eqnarray}
The last term of Eq.\eqref{V_osc} shows that $\phi$ interacts with other particles through the expansion of the conformal factor. Thus, if $V_{int}$ contains interactions of the Higgs boson with other particles,  the same interactions provide channels for the inflaton to decay into them and the corresponding decay rate is expected to be a function of the Higgs field value. Before the inflaton decays, the evolution of the universe is dominated by the mass term and thus the background equation of state is effectively $w_m=0$.

In the current case, the last term of \eqref{V_osc} is responsible for the Higgs-dependent inflaton decay. Following Ref.\cite{Ichikawa:2008ne}, one can generally take the following Higgs-dependent interactions:
\begin{eqnarray}\label{V_int^phi}
 V_{int}^{\phi} \supset y_a(h)\phi\bar\psi_a\psi_a + M_a(h)\phi\chi_a^2 + g_a(h)\phi^2\chi_a^2~,
\end{eqnarray}
where $\chi_a$ and $\psi_a$ are the scalar and spinor fields which constitute radiation in the early universe; the subscript $_a$ represents the species of particles. To achieve the modulated reheating scenario in our case, the coupling constants $y_a$, $M_a$, and $g_a$ must be functions of the Higgs field. Under this assumption the decay rate of the inflaton to the lowest order in coupling constants is given by
\begin{eqnarray}
 \Gamma(h) = \frac{y_a^2(h)}{8\pi}M_\phi + \frac{M_a^2(h)}{8\pi M_\phi} + \frac{g_a^2(h)}{16\pi M_\phi^3}\rho_\phi ~,
\end{eqnarray}
where the quadratic potential for $\phi$ has been applied.

Having obtained the decay rate of the inflaton, we now calculate the time when the inflaton field decays completely, and the universe enters the radiation dominant phase. In the following, we examine a specific example to illustrate the possibility of this decay process. We consider\footnote{\tc{We would like to remind that the real Higgs is a SU(2) doublet, ${\cal H}=(h_+,h_0)$, where $h_+$ and $h_0$ are the charged and neutral components, respectively. Since during inflation the energy scale is many order of magnitude larger than the vev of the Higgs, the symmetry breaking effect is irrelevant. Therefore, the leading term of $h$-dependent interaction is expected to be from the even Higgs operator ${\cal H}^\dagger{\cal H}\chi^2$. We thank the anonymous Referee for pointing out this important issue.}}
{\tc{
\begin{eqnarray}\label{V_int_h}
 V_{int}^\phi \supset \frac{\kappa}{M_p^2} h^2 {\chi^2\phi^2} ~,
\end{eqnarray}
}}
which could arise from the term $V_{int}\phi^2/M_p^2$ appearing in the last term of Eq.\eqref{V_osc}.
The corresponding decay rate is given by
{\tc{
\begin{eqnarray} \label{Gamma_h^2}
\Gamma(h) \simeq \frac{\kappa^2 h^4}{16\pi M_\phi^3M_p^4}\rho_\phi~.
\end{eqnarray}
}}
At the moment of the phase transition from inflaton domination to radiation domination, we have the condition $H\simeq \Gamma$. As we will show, this is the spatial hyper-surface where the primordial curvature perturbation can be calculated. At the reheating surface, $\rho_\phi\simeq 3M_p^2\Gamma^2$. By making use of Eq.\eqref{h_f}, we obtain the value of inflaton decay,
{\tc{
\begin{eqnarray}\label{phi_D}
 \phi_D \simeq \frac{\sqrt{3}\lambda^2\xi_{G}}{2\kappa^2\xi_{\Lambda}} M_p~,
\end{eqnarray}
}}
at the reheating surface. {\tc{ In order to connect the perturbative decay of inflaton with the inflationary phase smoothly, we expect $\phi_D\lesssim \phi_f$. As a consequence, it imposes an additional severe constraint, which requires the coefficient $\kappa$ needs to be finely tuned to satisfy $\kappa \gtrsim \lambda\sqrt{\xi_{G}/\xi_{\Lambda}}$. One may take into account the first and second terms in the interaction \eqref{V_int^phi} as well. The corresponding values of inflaton decay are estimated as $\frac{c_1^2\xi_\Lambda^2}{\lambda^2\xi_G^2}M_p$ and $\frac{c_2^2\xi_\Lambda}{\lambda^2\xi_G}M_p$, respectively, with $c_1$ and $c_2$ being the coefficients in front of these interaction terms. We can easily find both two values much smaller than the result obtained in \eqref{phi_D}. Therefore, one can conclude that the decay channel through the term \eqref{V_int_h} is generally dominant. }}

\section{Modulated reheating via Higgs boson}

In this section, we briefly review the standard calculations of the primordial power spectrum, the bispectrum and the trispectrum for the mechanism of modulated reheating, with the assumption that the inflaton decays on a spatial hypersurface with a variable local decay rate. Afterwards, we will apply this mechanism to our model in which the decay rate is a function of the Higgs boson and then study its cosmological implications.

\subsection{Field fluctuations during inflation}

We analyze primordial perturbations in a double-field inflation model involving kinetic couplings. We refer to \cite{GarciaBellido:1995qq, GrootNibbelink:2001qt, DiMarco:2002eb} for earlier studies of inflation models in terms of kinetically mixed double fields and \cite{Langlois:1999dw, Gordon:2000hv} for the paradigm of double-field inflation. Also, the topic of primordial perturbations in multiple-field inflation models was recently reviewed in \cite{Malik:2008im, Wang:2013zva}.

During inflation, it is convenient to decompose the field space of our model to directions parallel and orthogonal to the trajectory of background evolution. Along these two directions, one can define the adiabatic field, $\sigma$, and the entropy field, $s$, as follows,
\begin{eqnarray}
 \dot\sigma &=& \cos\theta \dot\phi + \Omega^{-1} \sin\theta \dot h~, \\
 \dot s &=& -\sin\theta \dot\phi +\Omega^{-1} \cos\theta \dot h~,
\end{eqnarray}
where the rotation angle is given by
\begin{eqnarray}\label{theta}
 \cos\theta = \frac{\dot\phi}{\sqrt{\dot\phi^2+\Omega^{-2}\dot h^2}}~,~
 \sin\theta=\frac{\Omega^{-1}\dot h}{\sqrt{\dot\phi^2+\Omega^{-2}\dot h^2}}~.
\end{eqnarray}

After that, we perturb the metric and fields up to linear order. One can introduce the field fluctuations along the adiabatic and entropy directions as follows,
\begin{eqnarray}
 \delta\sigma &=& \cos\theta\delta\phi + \Omega^{-1} \sin\theta \delta h~, \\
 \delta s &=& -\sin\theta \delta\phi +\Omega^{-1} \cos\theta \delta h~.
\end{eqnarray}
Neglecting interactions between these two modes, one can solve for the amplitude of entropy perturbation, $\delta s _*= H_*/2\pi$, during inflation and therefore the field fluctuation for the Higgs boson, which is given by
\begin{eqnarray}\label{delta_h_*}
 \delta{h}_* = \Omega_*\delta s _* = \Omega_*\frac{H_*}{2\pi}~,
\end{eqnarray}
where the subscript $_*$ denotes the moment of Hubble crossing.

\subsection{Higgs-modulated reheating}

The generation of curvature perturbation via Higgs-modulated reheating in a canonical model was recently suggested in \cite{DeSimone:2012qr, DeSimone:2012gq}, and soon it was pointed out in \cite{Choi:2012cp} that the Higgs-dependent interaction potential of the inflaton would be severely constrained by an upper bound on the value of the Higgs during inflation.

In the present paper, we extend the paradigm into the non-canonical model under consider. We show that this upper bound can be greatly relaxed by the relatively large value of the conformal factor and thus the corresponding parameter space is dramatically enlarged. Our scenario is easily extended to non-minimal inflation models.

In the treatment of local non-Gaussianity, the curvature perturbation can be expanded order by order as follows,
\begin{eqnarray}\label{zetalocal}
  \zeta(x) &=& \zeta_1(x)+\frac{3}{5}f_{\rm NL}\zeta_1^2(x)+\frac{9}{25}g_{\rm NL}\zeta_1^3(x)
  +{\cal O}(\zeta_1^4) \nonumber\\
  &=& \sum_{n=1}^{\infty}\frac{\zeta_n(x)}{n!}
  ~,
\end{eqnarray}
where $\zeta_1$ is the Gaussian fluctuation and $\zeta_n$ are the non-Gaussian components of order $\zeta_1^n$. The relation between $\zeta_n$ and the non-Gaussian parameters yields the following non-Gaussian estimators,
\begin{gather}\label{estimator}
  f_{\rm NL}=\frac{5}{6}\frac{\zeta_2}{\zeta_1^2}~,\qquad g_{\rm NL}=\frac{25}{54}\frac{\zeta_3}{\zeta_1^3}~.
\end{gather}
The correlation functions are defined as
\begin{gather}
  \langle \z{1}\z{2} \rangle = (2\pi)^3 P(k_1)\delta^3(\sum_{n=1}^2 \bk_n)~, \nonumber\\
  \langle \z{1}\z{2}\z{3} \rangle = (2\pi)^3 B(\bk_1,\bk_2,\bk_3)\delta^3(\sum_{n=1}^3 \bk_n)~,\nonumber\\
  \langle \z{1}\z{2}\z{3}\z{4} \rangle = (2\pi)^3 T(\bk_1,\bk_2,\bk_3,\bk_4)\delta^3(\sum_{n=1}^4 \bk_n)~,\label{def234}
\end{gather}
where $P(k_1)$ is related to the dimensionless power spectrum as
\begin{equation}
  {\cal P}_\zeta(k_1)\equiv \frac{k^3}{2\pi^2}P(k_1)~.
\end{equation}
Inserting the ansatz Eq.\eqref{zetalocal} into Eq.\eqref{def234}, one can relate the bispectrum $B$ and the trispectrum $T$ with $P$ as follows,
\begin{gather}
  \label{shapeB}
  B(\bk_1,\bk_2,\bk_3)=\frac{6}{5}f_{\rm NL}
  \left[P(k_1)P(k_2)+2{~\rm perm.}\right]~,
  \\
  \label{shapeT}
  T(\bk_1,\bk_2,\bk_3,\bk_4)=
  \frac{54}{25}g_{\rm NL}\left[ P(k_1)P(k_2)P(k_3) +3 {~\rm perm.} \right] \nonumber\\
  +\tau_{\rm NL} \left[ P(k_1)P(k_2)P(|\bk_1+\bk_3|)+11{~\rm perm.}\right]~.
\end{gather}
Note that if we neglect the nonlinear perturbations induced by self-interactions during inflation, which will be treated in the subsection \ref{subsec: NG}, there exists, in this case, a simple relation $\tau_{\rm NL}=(36/25)f^2_{\rm NL}$.

In the modulated reheating scenario, the decay of the inflaton occurs on a spatial hyper-surface with a varying local decay rate $\Gamma$, which is assumed to be a function of the Higgs boson in our model. Thus, the local Hubble parameter on the slice of modulated decay satisfies the condition $H=\Gamma(h)$\footnote{Within the framework of the multi-field inflationary cosmology, there exist many interesting scenarios for generating primordial curvature perturbation based on different choices of decay slices, such as the modulated curvaton decay mechanism \cite{Langlois:2013dh, Assadullahi:2013ey, Enomoto:2013qf} and the uniform curvaton decay mechanism \cite{Cai:2010rt}. All these scenarios are well established based on the validity of the $\delta N$ formalism.}. On super-Hubble scales, the curvature perturbation arisen from modulated decay can be written as
\begin{eqnarray}\label{zeta_123}
 \zeta_h(x) \simeq - \Theta_{1} \frac{\delta{h}}{h} - \frac{1}{2}\Theta_{2}\left(\frac{\delta{h}}{h}\right)^2
 - \frac{1}{6}\Theta_{3}\left(\frac{\delta{h}}{h}\right)^3 \bigg|_{D} ~,
\end{eqnarray}
where the subscript $_D$ denotes the moment of modulated decay. In our model the $\phi$ potential is dominated by its mass term after inflation, as shown in Eq.\eqref{V_osc}. We therefore obtain the coefficients as follows,
\begin{eqnarray}
 \Theta_{1} &=& \frac{h}{6}\frac{\Gamma_{,h}}{\Gamma}~, \\
 \Theta_{2} &=& \frac{h^2}{6}(\frac{\Gamma_{,hh}}{\Gamma} -\frac{\Gamma_{,h}^2}{\Gamma^2})~, \\
 \Theta_{3} &=& \frac{h^3}{6}(\frac{\Gamma_{,hhh}}{\Gamma} -3\frac{\Gamma_{,h}\Gamma_{,hh}}{\Gamma^2} +2\frac{\Gamma_{,h}^3}{\Gamma^3})~.
\end{eqnarray}

At linear order in the curvature perturbation, the coefficient $\Theta_1$ is typically of the order $O(1)$ and thus $\zeta$ is mainly determined by $\delta{h}/h$ at the moment of modulated decay. In the conventional scenario of modulated reheating, one can approximately take $\delta{h}_D$ to be the amplitude of entropy field at the moment of Hubble-crossing during inflation. However for the model of AS inflation, the Higgs boson and the inflaton are coupled through a conformal factor in front of its kinetic term. Making use of Eq.\eqref{delta_h_*}, one finds,
\begin{eqnarray}\label{delta_h_D}
 \delta{h}_D = \frac{\Omega_D}{\Omega_*} \delta{h}_* = \frac{\Omega_DH_*}{2\pi} ~.
\end{eqnarray}
Moreover, the value of the Higgs at the slice of modulated decay can be related to the Hubble-crossing value by introducing a general function
\begin{eqnarray}
 h_D = g(h_*)~,
\end{eqnarray}
where its detailed form is determined by the explicit potential of the entropy field. For example, in the model under consider, $g(h_*)\propto h_* \sim h_I$. For simplicity, we assume that $g(h_*)$ is linear.

As a consequence, the power spectrum of curvature perturbation due to modulated reheating is given by
\begin{eqnarray}
 P_{\zeta_h} = \Theta_1^2\frac{\delta{h}_D^2}{h_D^2}
 \simeq \Theta_1^2 \Omega_D^2 \frac{H_*^2}{4\pi^2 h_*^2}~,
\end{eqnarray}
where we have applied the field fluctuation Eq.\eqref{delta_h_D} to obtain the second, approximate equality, expression.

\subsection{Observables at linear order}

If we further neglect the variation of the Hubble parameter during inflation, then we can obtain an approximate power spectrum from Higgs-modulated reheating,
\begin{eqnarray}
 P_{\zeta_h} \simeq \frac{\lambda}{8\pi^2} \Theta_1^2 \frac{\Omega_D^2}{\Omega_I^4}~,
\end{eqnarray}
by inserting the approximate relation Eq.\eqref{h_i}. Note that the usual decay rate is a power-law function of the Higgs boson such as that in Eq.\eqref{Gamma_h^2}, considered in the previous section. Thus $\Theta_1^2$ is typically of the order $O(0.01 \sim 1)$. The coefficient $\Omega_D$ is totally determined by $\phi_D$ as provided by Eq.\eqref{phi_D}, and it therefore depends only on $\lambda$ and $\kappa$; numerically $\Omega_D$ is of the order $O(1)$. Finally, we note that the power spectrum generated from the Higgs-modulated reheating is determined by the Higgs coupling $\lambda$, the interaction coupling $\kappa$, and the conformal factor during inflation $\Omega_I$ (or, equivalently, the value of inflaton $\phi_I$).

In addition to the curvature perturbation generated by modulated reheating, there exists the intrinsic curvature perturbation due to the inflaton fluctuation, which takes the form:
\begin{eqnarray}
 P_{\zeta_\phi} = \frac{H_I^2}{8\pi^2\epsilon M_p^2}~.
\end{eqnarray}
It is convenient to define a Higgs-to-curvature ratio
\begin{eqnarray}\label{q_h}
 q_h \equiv \frac{P_{\zeta_h}}{P_{\zeta_h}+P_{\zeta_\phi}}
 =\frac{\epsilon\lambda\Theta_1^2\Omega_D^2 M_p^2}{\epsilon\lambda\Theta_1^2\Omega_D^2 M_p^2 +\Omega_I^4 H_I^2} ~,
\end{eqnarray}
to characterize the relative contribution of Higgs fluctuations. If the main contribution to generating primordial curvature perturbation is due to the modulated reheating, then we expect $q_h \simeq 1$. By choosing a group of values for the model parameters such as that provided in Fig.\ref{Graph_bg} and the decay rate given by Eq.\eqref{Gamma_h^2}, one finds: $\epsilon \sim 10^{-4}$, $\Omega_I \sim 10$, $\Omega_D \sim 1$, $\Theta_1^2 \sim 0.1$ and $\lambda \sim 0.13$. Under this particular parameter choice, we find that the mechanism of Higgs modulated reheating dominates as long as $H_I^2 < 10^{-9}M_p^2$ without any fine-tuning.

These two power spectra actually show different signatures on their spectral indices. Specifically, their spectral indices are given by
\begin{eqnarray}
 n_{\zeta_\phi}-1 &=& -6\epsilon +2\eta_\phi = -6\epsilon +\frac{2U_{,\phi\phi}}{3H^2}~, \\
 n_{\zeta_h}-1 &=& -2\epsilon +2\eta_h = -2\epsilon +\frac{2V_{,hh}}{3\Omega_I^4H^2}~,
\end{eqnarray}
which are calculated at the moment of Hubble-crossing. In addition, the primordial tensor perturbations are only dependent on the inflationary Hubble parameter, whose spectrum is given by
\begin{eqnarray}
 P_T = \frac{2H^2}{\pi^2M_p^2}~,
\end{eqnarray}
As usual, the spectral tilt is given by
\begin{eqnarray}
 n_T = -2\epsilon~.
\end{eqnarray}
In the modulated reheating scenario the conventional tensor-to-scalar ratio $r_T$ is now defined as
\begin{eqnarray}
 r_T \equiv \frac{P_T}{P_{\zeta_h}+P_{\zeta_\phi}} = 16\epsilon (1-q_h)~,
\end{eqnarray}
which indicates that the amplitude of primordial gravitational wave is doubly suppressed in the Higgs-modulated reheating mechanism since both $\epsilon$ and $1-q_h$ are small quantities.

\subsection{Non-Gaussianities}
\label{subsec: NG}

In contrast to the prediction of a canonical single-field inflation model \cite{Maldacena:2002vr}, a salient feature of the modulated reheating mechanism is that sizable amplitudes of primordial non-Gaussianities can be obtained under suitable parameter choices. In this subsection we study the curvature perturbation beyond linear level. As we expected, the curvature perturbation is mainly sourced by the Higgs fluctuations. For the time being, we ignore the nonlinear effects of inflaton, which are generally suppressed by slow-roll parameters. For the nonlinear fluctuations seeded by the Higgs fluctuations, there exist two categories of seeds, with one being proportional to the connected correlators of the Higgs and the other being an intrinsically non-Gaussian distribution \cite{DeSimone:2012gq}.

\subsubsection{Non-Gaussianities from modulated reheating}

The first type of non-Gaussianities originates from the field fluctuations at super-Hubble scales during the process of post-inflation modulated reheating. In this era the Higgs field is considered Gaussian while the non-Gaussianity is induced by the nonlinear conversion from $\delta h$ to $\zeta$. One can insert the second and third order curvature perturbations in Eq.\eqref{zeta_123} into the non-Gaussian estimator Eq.\eqref{estimator} and obtain this part of the ``universal" nonlinearity parameters:
\begin{eqnarray}
\label{f_NL^un}
 f_{\rm NL, un}^{\rm local} &=& 5 q_h^2 \left( 1-\frac{\Gamma\Gamma_{,hh}}{\Gamma_{,h}^2} \right)~,\\
\label{g_NL^un}
 g_{\rm NL, un}^{\rm local} &=& \frac{50}{3} q_h^3 \left( 2 - 3\frac{\Gamma\Gamma_{,hh}}{\Gamma_{,h}^2} +\frac{\Gamma^2\Gamma_{,hhh}}{\Gamma_{,h}^3}  \right)~,
\end{eqnarray}
which are of local type.

{\tc{In particular, for the interaction term considered in \eqref{V_int_h}, the decay rate is proportional to $h^4$ and $q_h \simeq 1$ can be obtained under a reasonable set of values of model parameters. As a consequence, one obtains $f_{\rm NL, un}^{\rm local} \simeq 5/4$ and $g_{\rm NL, un}^{\rm local} \simeq 25/12$. These nonlinear parameters are sizable when compared with those in slow-roll inflation models, but the corresponding non-Gaussianities are still difficult to test observationally.}}

\subsubsection{Non-Gaussianities from Higgs Self-interaction during inflation}

The second type of non-Gaussianities originates from the non-quadratic potential of the lighter field, which in our model corresponds to the Higgs potential: $V(h)\simeq \lambda h^4/4$. In fact, this self-interaction of the scalar field can also generate primordial non-Gaussian fluctuations during inflation.

Following \cite{DeSimone:2012gq} (see also \cite{Zaldarriaga:2003my}), the $n$-point correlation function of $\delta h$ is evaluated by
\begin{widetext}
\begin{eqnarray}
 \langle \delta h_{{\bf k}_1}(\tau) \delta h_{{\bf k}_2}(\tau) \cdot\cdot\cdot \delta  h_{{\bf k}_n}(\tau) \rangle =
 -i \langle | \int_{-\infty}^\tau a d\tau^\prime [ \delta h_{{\bf k}_1}(\tau) \delta h_{{\bf k}_2}(\tau) \cdot\cdot\cdot \delta  h_{{\bf k}_n}(\tau), ~ H^{(n)}_{int}(h(\tau^\prime) ] | \rangle~,
\end{eqnarray}
\end{widetext}
where $H_{int}^{(n)}$ is the $n$-th order interaction Hamiltonian. Here by $n$-th order we mean the part of $H_{int}$ that is of the order ${\cal O}(\delta h^n)$. In our model the Higgs field is conformally coupled to the inflaton due to the RG running gravitational constant. The corresponding field fluctuation is expressed as
\begin{eqnarray}
\delta h_{\bf k}&=&\frac{i\Omega H}{\sqrt{2k^{3}}}(1+ik\tau)e^{-ik\tau}~,
\end{eqnarray}
during inflation. In addition, the interaction Hamiltonian of the Higgs field takes the form
\begin{eqnarray}
H^{(n)}_{int}(\tau)&=&\int d^3x a^3{\cal H}^{(n)}_{int}~\nonumber\\
&=&\int d^3x a^3\Omega^{-4}(\phi)\frac{1}{n!}V^{(n)}(h)\delta h^n~,
\end{eqnarray}
where $V^{(n)}(h)\equiv\partial^nV/\partial h^n$ is the $n$-th derivative of the potential $V(h)$ with respect to the field $h$, and $V(h)$ as well as $\Omega(\phi)$ are given by Eq.(\ref{V_h}) and Eq.(\ref{Omega}), respectively.

We perform the integrals appearing in the correlation functions and find
\begin{eqnarray}
 \langle \delta h_{{\bf k}_1} \delta h_{{\bf k}_2} \cdot\cdot\cdot \delta  h_{{\bf k}_n} \rangle &=&
 \frac{ ( \Omega_\ast H_\ast)^{2n-4} V_\ast^{(n)} K^3}{\prod^n_{i=1}(2k_i^3)} \delta^3(\sum^n_{i=1}{\bf k}_i)
  \nonumber\\
 && \times (2\pi)^3  I_n({\bf k}_1,{\bf k}_2,\cdot\cdot\cdot,{\bf k}_n) ~,
 \nonumber\\
\end{eqnarray}
where a kernel integral function has been introduced as follows,
\begin{equation}
 I_n \equiv 2 \text{Re} \Big[ -i\int^{\tau_{end}}_{-\infty} \frac{d\tau^\prime}{K^3\tau^{\prime 4}}
 \prod^n_{i=1} (1-ik_i\tau) e^{iK\tau} \Big] ~,
\end{equation}
with $K$ defined as $K\equiv\sum^n_{i=1}k_i$.

For 3- and 4-point correlation functions which are of observable interests, we identify, according to the definitions in Eq.(\ref{def234}), the following expressions:
\begin{eqnarray}
\label{Bnun}
 B^{n-un}_{\delta h}({\bf k}_1,{\bf k}_2,{\bf k}_3)
 &=& \frac{\Omega_\ast^2H_\ast^{2}V_\ast^{(3)}K^3}{\prod^3_{i=1}(2k_i^3)} I_3 \nonumber\\
 &=& \frac{3\lambda h_\ast\Omega_\ast^2H_\ast^{2}K^3}{4\prod^n_{i=1}k_i^3}
 \Big[\frac{8}{9}-\frac{2\sum_{i<j}k_ik_j}{K^2} \nonumber\\
 &&-\frac{2}{3}(\gamma_E+N_{K})\frac{\sum_ik_i^3}{K^3}\Big] ~,
\end{eqnarray}
and
\begin{eqnarray}
\label{Tnun}
 T^{n-un}_{\delta h}({\bf k}_1,{\bf k}_2,{\bf k}_3,{\bf k}_4)
 &=& \frac{\Omega_\ast^4H_\ast^{4}V_\ast^{(4)}K^3}{\prod^4_{i=1}(2k_i^3)}I_4 \nonumber\\
 &=&\frac{3\lambda\Omega_\ast^4H_\ast^{4}K^3}{8\prod^n_{i=1}k_i^3}
 \Big[\frac{8}{9}-\frac{2\sum_{i<j}k_ik_j}{K^2}\nonumber\\
 && +2\frac{\prod_ik_i}{K^4}  -\frac{2}{3}(\gamma_E+N_K)\frac{\sum_ik_i^3}{K^3}\Big] ~, \nonumber\\
\end{eqnarray}
where $\gamma_E\simeq 0.58$ is the Euler-Masheroni constant and $N_K$ is the e-folding number for the perturbation mode with a fixed $K$ crossing the Hubble radius until the end of inflation $\tau_{end}$. As introduced in previous section, the subscript ``$\ast$" indicates the values at Hubble crossing.

We first calculate the non-Gaussianities of {\it equilateral} type. One can estimate the correlation functions $B^{n-un}_{\delta h}$ and $T^{n-un}_{\delta h}$  under the particular limit that all the $k_i$'s are of the same value. As a result,  substituting Eqs. \eqref{Bnun}, \eqref{Tnun} into the expressions Eq.\eqref{def234} yields the nonlinearity parameters of equilateral type as follows,
\begin{eqnarray}
\label{f_NLint^e}
 f_{\rm NL, int}^{\rm equil} &\simeq& - \frac{5 \lambda h_\ast^2}{3 \Theta_1 \Omega_\ast^2 H_\ast^2} q_h^2 \bigg( N_K+\gamma_E-3 \bigg) ~,\\
\label{g_NLint^e}
 g_{\rm NL, int}^{\rm equil} &\simeq& - \frac{25 \lambda h_\ast^2}{27 \Theta_1^2 \Omega_\ast^2 H_\ast^2} q_h^3
 \bigg( N_K+\gamma_E-\frac{169}{48} \bigg) ~.
\end{eqnarray}

Next we study the non-Gaussianities originated from the self-interaction of the Higgs field during inflation in the squeezed limit where we assume $k_1\ll k_2, k_3$ (for bispectrum) and $k_1\ll k_2, k_3, k_4$ (for trispectrum). The same scenario in the framework of GR was discussed in \cite{DeSimone:2012gq}. Here we directly calculate the correlation function of the curvature perturbation and then derive the nonlinearity parameters:
\begin{eqnarray}
\label{f_NLint^l}
 f_{\rm NL, int}^{\rm local} &\simeq& - \frac{5 \lambda h_\ast^2}{3 \Theta_1 \Omega_\ast^2 H_\ast^2} q_h^2 \bigg( N_K+\gamma_E-\frac{7}{3} \bigg) ~,\\
\label{g_NLint^l}
 g_{\rm NL, int}^{\rm local} &\simeq& - \frac{25 \lambda h_\ast^2}{27 \Theta_1^2 \Omega_\ast^2 H_\ast^2} q_h^3
 \bigg( N_K+\gamma_E-3 \bigg) ~.
\end{eqnarray}

From the above results, we can immediately see that the primordial non-Gaussianities due to the self-interaction of the Higgs field during inflation are negative. This is a novel feature in the Higgs modulated reheating scenario. A similar feature was observed in \cite{DeSimone:2012gq} in the framework of standard GR, but in our model the amplitude of nonlinearity parameters involve a new parameter which is the conformal factor $\Omega$.

Consider for the moment the primordial curvature perturbation due solely to the modulated reheating, $q_h \simeq 1$. By choosing a group of reasonable values for the model parameters such as that provided in the previous section, we find $\epsilon \sim 10^{-4}$, $\Omega_I \sim 10$, $\Theta_1^2 \sim 0.1$ and $\lambda \sim 0.13$. In addition, there is a theoretical lower bound: $h_I>2 H_I/\sqrt{\lambda}$. By assuming $N_K\sim 50$, one obtains $f_{\rm NL, int}^{\rm local}\lesssim -10$. We see that this particular parameter choice appears to be incompatible with the newly released Planck data. This points to the necessity of performing an analysis of the observational constraints on our model. This is the main content of the next section.

\section{Constraint on model parameters by Planck}

Recently the Planck mission has released data on CMB anisotropy. The results highly constrain cosmological parameters with unprecedented accuracy. Specifically, the amplitude and spectral index of primordial curvature perturbation are determined to be $10^9 P_\zeta = 2.23 \pm 0.16$, $n_s = 0.9603 \pm 0.0073$ ($68\%$ C.L.) at the pivot scale $k=0.002 {\rm Mpc}^{-1}$ \cite{Ade:2013lta}. Moreover, there is no significant evidence for primordial curvature perturbation deviating from Gaussian distribution. In particular, the bounds on nonlinearity parameters are quoted as: $f_{\rm NL}^{\rm local} = 2.7 \pm5.8$, $f_{\rm NL}^{\rm equal} = -42 \pm 75$ ($68\%$ C.L.) \cite{Ade:2013tta}. In addition, the upper bound on the tensor-to-scalar ratio is given by $r_T<0.11$ at $2\sigma$ level.

In our model of RG improved Higgs modulated reheating, there are eight model parameters. Among these parameters, $H_\ast$, $\Omega_\ast$ and $\epsilon$ are associated with the background model; $h_\ast$ and $\lambda$ are related to the details of the Higgs model; and $\Theta_1$, $\Theta_2$ and $\Theta_3$ are determined by specific forms of the decay process, respectively. To be more explicit, $\lambda$ is basically constrained by particle physics experiments such as those at LHC, which have determined that $\lambda \simeq 0.13$. At present the Planck data has not yet imposed strong constraints on the tri-spectrum and thus $\Theta_3$ is free. We therefore only need to analyze the combined constraints on the remaining parameters.


\begin{figure}[htbp]
\includegraphics[scale=0.4]{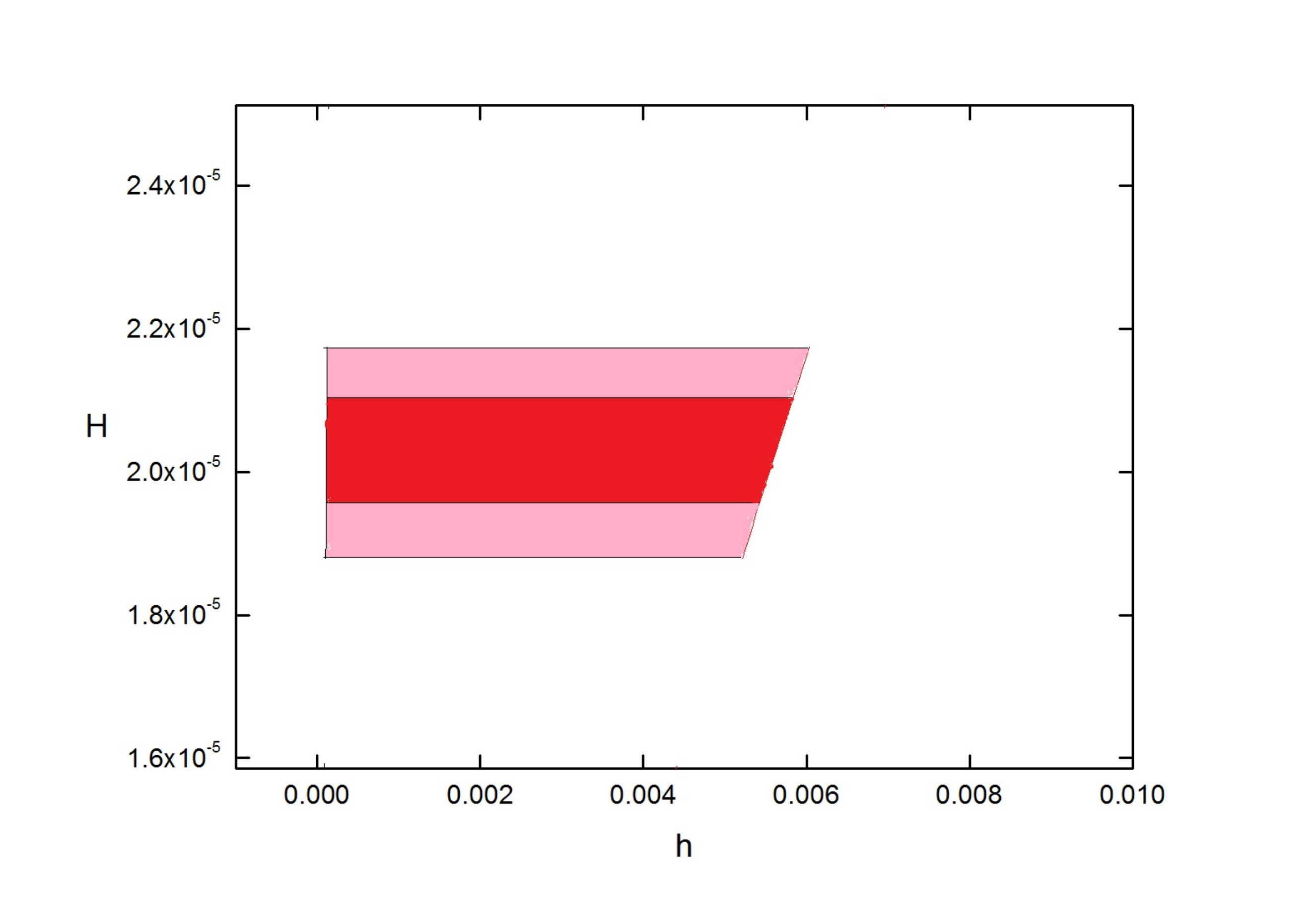}
\caption{Observational and theoretical constraints on $h_\ast$ and $H_\ast$ for $q_h=0$. The viable parameter space is within the red region (C.L. $68\%$) and the light red region (C.L. $95\%$).}
\label{fig:conq1}
\end{figure}

\begin{figure}[htbp]
\includegraphics[scale=0.4]{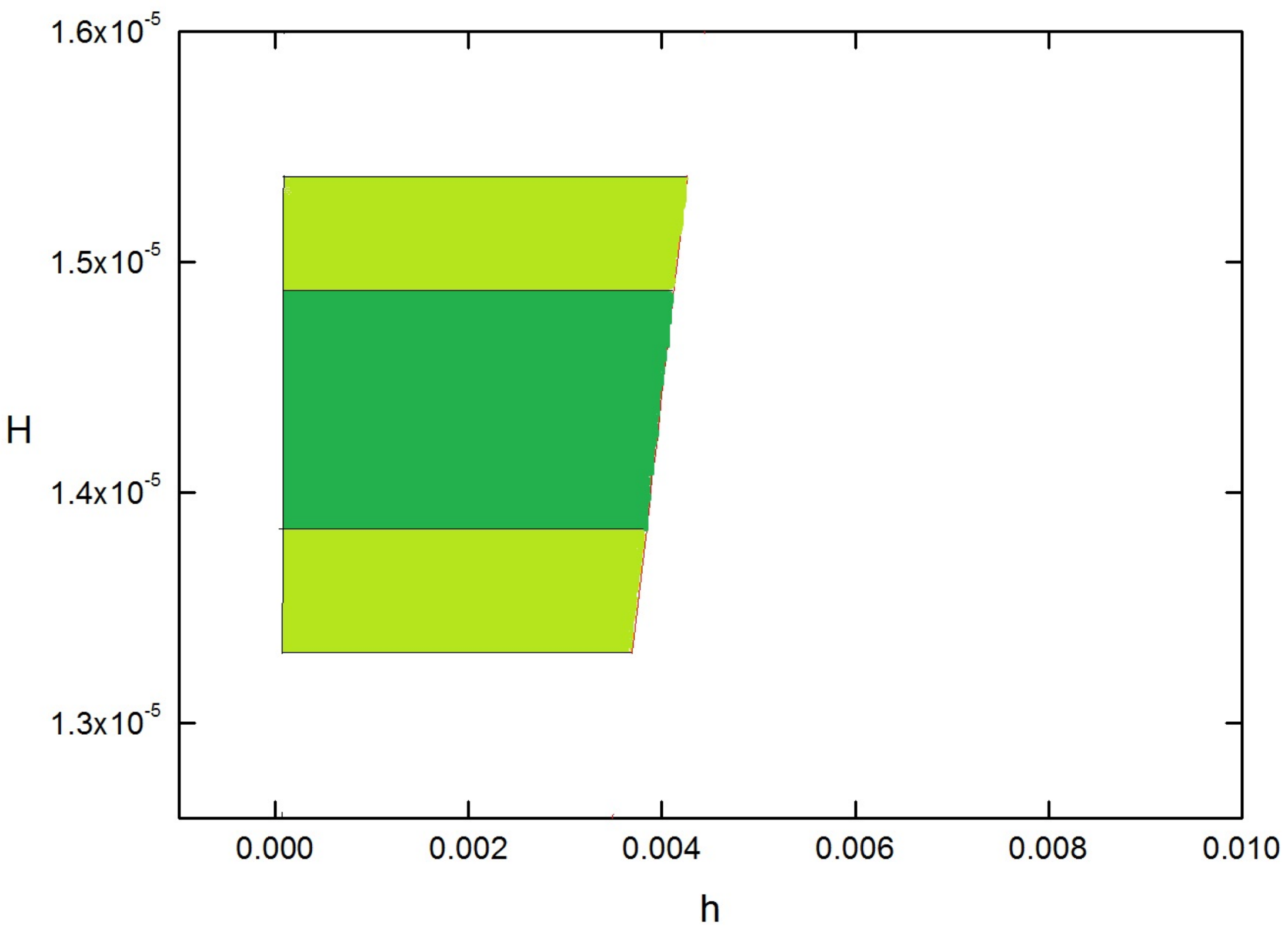}
\caption{Observational and theoretical constraints on $h_\ast$ and $H_\ast$ for $q_h=0.5$. The viable parameter space is within the green region (C.L. $68\%$) and the light green region (C.L. $95\%$). }
\label{fig:conq2}
\end{figure}

\begin{figure}[htbp]
\includegraphics[scale=0.4]{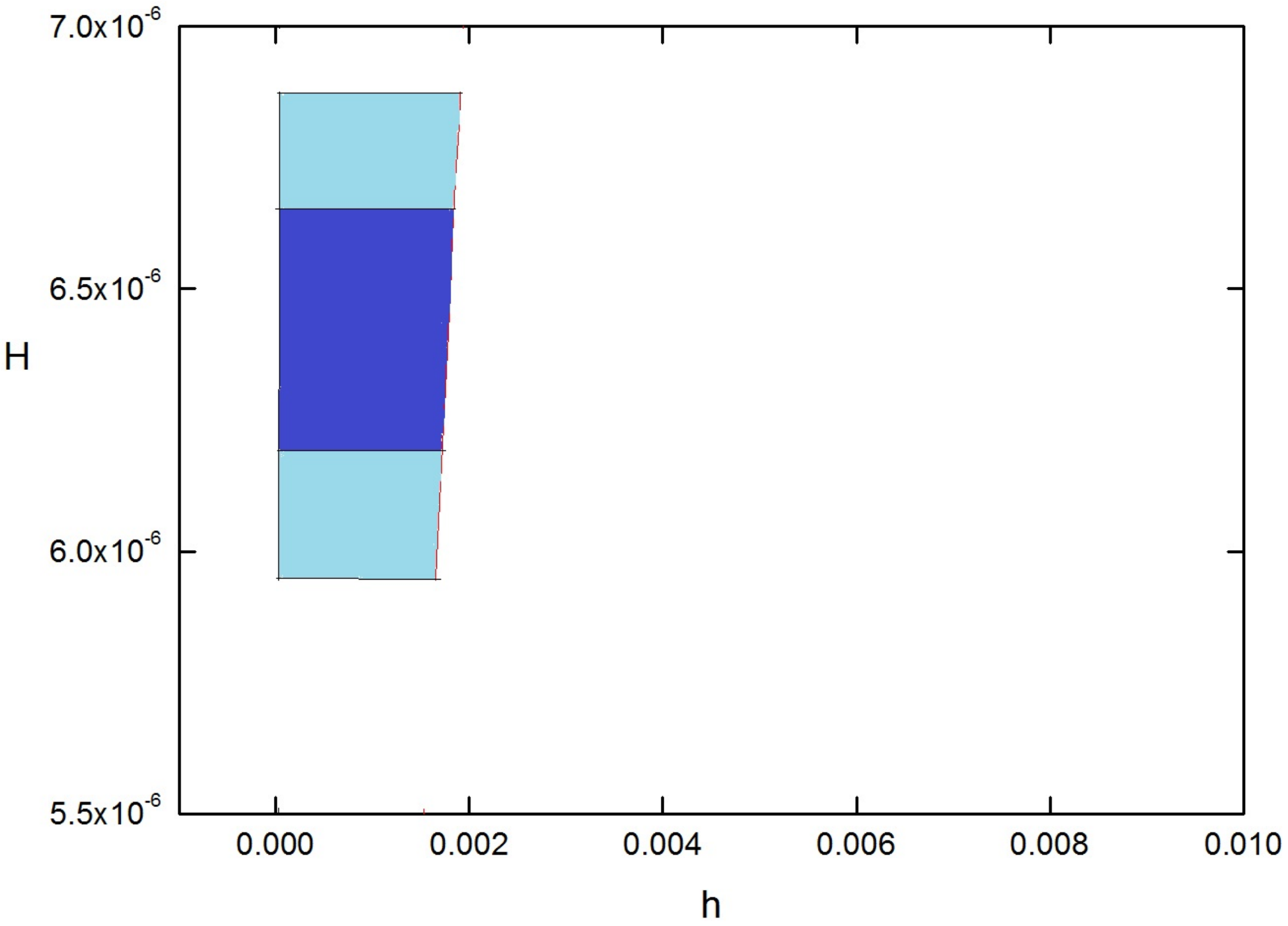}
\caption{Observational and theoretical constraints on $h_\ast$ and $H_\ast$ for $q_h=0.9$. The viable parameter space is within the blue region (C.L. $68\%$) and the light blue region (C.L. $95\%$). }
\label{fig:conq3}
\end{figure}

Allowing all plausible values for $\epsilon$ and $\Theta_1$, we obtain a combined constraint on the inflationary Hubble rate $H_\ast$ and the amplitude of the Higgs field $h_\ast$ for different choices of $q_h$ as depicted in Figs. \ref{fig:conq1}, \ref{fig:conq2} and \ref{fig:conq3}. One can read from the figures that $H_\ast$ is constrained to be of the order $O(10^{-5} \thicksim 10^{-6})M_p$ and $h_\ast$ is allowed to vary between $10^{-4}M_p$ and $10^{-2}M_p$ during inflation. One can also see from the figures that the larger $q_h$ is, the smaller the central value of $H$ will be, consistent with the definition of $q_h$, namely Eq. (\ref{q_h}). This indicates that the more Higgs contributes to the final power spectrum, the lower scale inflation we can get.

\section{Conclusion}

In this paper, we studied the RG improved inflationary cosmology where the decay rate of inflaton is modulated by a second scalar field, which we identified as a Higgs. In this model the background evolution is driven by a RG running cosmological constant and gravitational constant with their RG flows satisfying the AS behavior. By choosing Einstein-Hilbert truncation, we find this model is classically equivalent to a model of $R^2$ gravity. The elegant property of this model is that it can give rise to a sufficiently long inflationary phase at high energy scale and smoothly exit to standard GR after inflation. Moreover, an RG running gravitational constant can assist a second scalar field to vary slowly without an extremely flat potential since the slow-roll parameters associated with this field are greatly suppressed by a large value of the conformal factor. As a consequence, this scalar field seeds isocurvature perturbations during inflation which can be converted into primordial curvature perturbation under a suitable mechanism. We consider this mechanism as the process of modulated reheating.

Based on this mechanism, we performed a detailed analysis on the power spectrum and non-Gaussianities of primordial cosmological perturbations. We then confronted our model with the recently released Planck data and concluded that a viable parameter space exists, although it is highly constrained.
Although this model suffers from the fine-tuning problem, the scenario under present study points to a new possible connection between particle physics and early universe cosmology.

We conclude by mentioning that the mechanism of Higgs modulated reheating can be generalized to an arbitrary non-minimal inflationary model or a model of $f(R)$ inflation. By relaxing theoretical requirements of AS gravity, the parameter space available to such a mechanism is increased. This topic is of phenomenological interest.

\section*{Acknowledgments}
It is a pleasure to thank Robert Brandenberger, Keisuke Izumi, Andrea De Simone, Antonio Riotto, Teruaki Suyama, and Yi Wang for useful discussion and comments on the manuscript. The work of YFC is supported in part by Department of Physics in McGill University. YCC, PC, and TQ are supported by Taiwan National Science Council under Project No. NSC 101-2923-M-002-006-MY3 and 101-2628-M-002-006- and by Taiwan National Center for Theoretical Sciences (NCTS). PC is in addition supported by US Department of Energy under Contract No. DE- AC03-76SF00515. DAE is supported in part by the Cosmology Initiative at ASU and by DOE grant DE-SC0008016. YFC is very grateful to PC and TQ for their hospitality during his visit to the Leung Center for Cosmology and Particle Astrophysics at National Taiwan University while this work was initiated.

\end{document}